\documentclass[%
 reprint,%
 amssymb, amsmath,%
 aip,apl,%
]{revtex4-1}

\usepackage{docs}%
\usepackage{bm}%
\usepackage[colorlinks=true,linkcolor=blue]{hyperref}%
\expandafter\ifx\csname package@font\endcsname\relax\else
 \expandafter\expandafter
 \expandafter\usepackage
 \expandafter\expandafter
 \expandafter{\csname package@font\endcsname}%
\fi
\usepackage{dcolumn}   
\usepackage{bm}        
\usepackage{amssymb}   
\usepackage{color}

\newcommand{\figref}[1]{Fig.~\ref{#1}}

\usepackage{color}

\usepackage{amsmath}
\usepackage{verbatim} 
\usepackage{color}
\usepackage[section] {placeins}
\usepackage{graphicx}  
\usepackage{epstopdf}

\usepackage[caption=false,font=footnotesize]{subfig}

\begin{document}

\title{Photonic-Band-Gap Gyrotron Amplifier with Picosecond Pulses}%

\author{Emilio A. Nanni}%
\email{nanni@slac.stanford.edu}
\affiliation{Plasma Science and Fusion Center, Massachusetts Institute of Technology, 77 Mass. Ave.,Cambridge, MA 02138, USA}%
\affiliation{SLAC National Accelerator Laboratory, Stanford University, 2575 Sand Hill Road, Menlo Park, CA 94025, USA}%

\author{Sudheer Jawla }%
\affiliation{Plasma Science and Fusion Center, Massachusetts Institute of Technology, 77 Mass. Ave.,Cambridge, MA 02138, USA}%

\author{Samantha M. Lewis}%
\affiliation{Plasma Science and Fusion Center, Massachusetts Institute of Technology, 77 Mass. Ave.,Cambridge, MA 02138, USA}%

\author{Michael A. Shapiro}%
\affiliation{Plasma Science and Fusion Center, Massachusetts Institute of Technology, 77 Mass. Ave.,Cambridge, MA 02138, USA}%

\author{Richard J. Temkin}%
\affiliation{Plasma Science and Fusion Center, Massachusetts Institute of Technology, 77 Mass. Ave.,Cambridge, MA 02138, USA}%

\date{\today}%

\begin{abstract}
{We report the amplification of 250~GHz pulses as short as 260~ps without observation of pulse broadening using a photonic-band-gap circuit gyrotron traveling-wave-amplifier. The gyrotron amplifier operates with 38~dB of device gain and 8~GHz of instantaneous bandwidth. The operational bandwidth of the amplifier can be tuned over 16 GHz by adjusting the operating voltage of the electron beam and the magnetic field. The amplifier uses a 30~cm long photonic-band-gap interaction circuit to confine the desired TE$_{03}$-like operating mode while suppressing lower order modes which can result in undesired oscillations. The circuit gain is $>$55~dB for a beam voltage of 23~kV and a current of 700~mA. These results demonstrate the wide bandwidths and high gain achievable with gyrotron amplifiers. The amplification of picosecond pulses of variable lengths, 260-800~ps, shows good agreement with theory using the coupled dispersion relation and the gain-spectrum of the amplifier as measured with quasi-CW input pulses.}
\end{abstract}

\maketitle


Amplification of picosecond pulses to high powers in the millimeter and sub-millimeter range is of interest for a wide variety of applications including radar, communications, particle accelerators and spectroscopy. Many of these applications require the source to provide phase stability and control and in all cases a wide gain-spectrum allows for signal preservation. 
Among the most exciting applications for high power THz sources is pulsed dynamic nuclear polarization (DNP) nuclear magnetic resonance (NMR) spectroscopy. Microwave-driven DNP experiments are now recognized as a powerful method of enhancing signals in solid state and solution NMR and imaging. DNP improves the sensitivity of NMR spectra by about a factor of more than 100.\cite{barnes2008} DNP has been demonstrated successfully up to 16.4~T ($^1$H/e$^-$ frequencies of 700~MHz / 460~GHz) using a gyrotron oscillator.\cite{barnes2012dynamic} However, modern NMR spectroscopy is moving to higher magnetic fields, where much better spectral resolution is achieved. This increase in magnetic field results in a decrease in the achievable enhancement as $\omega_0^{-1}$ for the cross effect and $\omega_0^{-2}$ for the solid effect,\cite{Goldman1970} where $\omega_0$ is the frequency of operation. Circumventing these reductions in DNP enhancements requires the use of time domain-DNP (pulsed-DNP) experiments. The extension of pulsed-DNP to higher magnetic fields requires the development of THz sources at the relevant frequencies with phase control and the ability to generate the required pulse sequences. Gyrotron amplifiers are the most promising technology available because they offer the possibility of achieving high gain and broad bandwidth even as the frequency of operation moves into the THz band.

Techniques for generating ultra-short pulses have been recently proposed \cite{ginzburg2015mechanisms,ginzburg2017generation} using helical gyro-TWTs with large gain bandwidth.\cite{denisov1998gyrotron,he2011w} Pulses as short as one nanosecond have been generated in a THz gyrotron oscillator operating in a Q-switched regime.\cite{alberti2013nanosecond} Short pulse amplification has been reported in our earlier experiment \cite{kim2010prl} where we presented an experimental study of amplification of pulse broadening of sub-nanosecond pulses at 140~GHz. Pulses as short as one nanosecond showed non-distorted amplification of such pulses in a gyro-TWT amplifier. In that experiment, broadening for pulses below one nanosecond was observed due to the frequency dependence of the group velocity near cutoff and gain narrowing by the finite gain bandwidth of the amplifier. In the present experiment at 250~GHz, based on a photonic-band-gap (PBG) waveguide gyrotron amplifier, 400~ps pulses were observed with no measurable distortion to the temporal profile of the pulse. In addition, we have made the first observation of pulse shortening, with compression $\sim40\%$ in temporal duration.

A gyrotron amplifier operates by the resonant interaction between the eigenmodes of an interaction circuit (waveguide), typically cylindrical, and a mildly relativistic electron beam that is gyrating in a constant axial magnetic field. The amplifier is operated under conditions that suppress self-start oscillations, including backward wave oscillations that could disrupt the operation of the device. Amplification is achieved in a gyrotron amplifier by a convective instability that results from the interaction of a mildly relativistic, annular, gyrating electron beam and a transverse electric (TE) mode in a waveguide immersed in a strong static axial magnetic field $(B_0)$. The grazing intersection between the dispersion relation lines of the cyclotron resonance and a TE waveguide mode near the waveguide cutoff results in high gain and moderate bandwidth. The Doppler shifted electron beam resonance condition is given by
\begin{equation}
\omega - {n\Omega}/{\gamma} - k_z v_z = 0
\label{eq:beamline}
\end{equation}
and the waveguide mode dispersion relation is 
 \begin{equation}
\omega^2 - k_z^2 c^2 - k_\perp^2c^2 = 0,
\label{eq:waveline}
\end{equation}
where $\omega$  is the frequency of the wave;   $\Omega=eB_0/m_e$ is the non-relativistic cyclotron frequency of the gyrating electrons; $e$ and $m_e$ are, respectively, the charge and the rest mass of the electron;  $\gamma$ is the relativistic mass factor; $n=1$ is the cyclotron harmonic number; $k_z$ and $k_\perp$  are the longitudinal and transverse propagation constants, respectively, of the waveguide mode; $v_z$ is the axial velocity of the electrons and  $c$ is the speed of light. This requires the design of an interaction circuit that supports the propagation of a mode at the correct frequency while appearing lossy to frequencies and modes outside the region of interest. This is usually performed by inserting lossy ceramics, including severs and selecting geometries with favorable dispersion relations.\cite{joye2009demonstration} The gyrotron amplifier used in this experiment relies on a PBG interaction circuit composed of an array of metal rods to support the desired mode of operation while rejecting spurious modes.\cite{nanni2013photonic} The PBG gyrotron amplifier demonstrated record gain, power and bandwidth for a device operating in this frequency range. For this experiment, it was upgraded to include a Vlasov launcher for the input coupler to maximize the operational bandwidth of the circuit. 

The experimental setup and optimized performance of the PBG gyrotron amplifier were previously published.\cite{nanni2013photonic,nanni2013high} During these experiments it was observed that both the gain and bandwidth of the device were limited by the performance of the wrap-around input coupler which excites the TE$_{03}$-like mode in the interaction circuit of the amplifier. In order to improve the device performance it was determined that the use of a Vlasov launcher on the input (and output) of the device could provide for a greatly improved gain-bandwidth. In addition, the interaction circuit length was increased from 26 cm to 30 cm to prevent the excitation of oscillations in the uptapers of the Vlasov launcher which had been previously observed. By increasing the circuit length the uptapers of both launchers are located outside the flat-top of the magnetic field.

\begin{figure}[!t]
  \centering
  \includegraphics[width=0.38\textwidth]{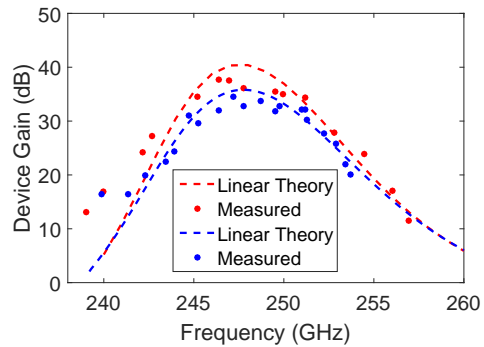}
	\caption {A comparison between the linear theory\cite{nusinovich1992theory} (dashed lines) including insertion losses\cite{nanni2016amplification} and the measured (dots) gain-bandwidth of the amplifier. The red line and dots correspond to the nominal operating conditions of 22.8~kV, 675~mA, $\alpha$=0.45 and $B_0$=8.90~T (Operating Point 1).  A second operating point is shown with a blue line and dots for 25.2~kV, 511~mA, $\alpha$=0.5 and $B_0$=8.90~T (Operating Point 2). }
  \label{fig:gainbw}
\end{figure}

\begin{figure}[!t]
  \centering
  \includegraphics[width=0.38\textwidth]{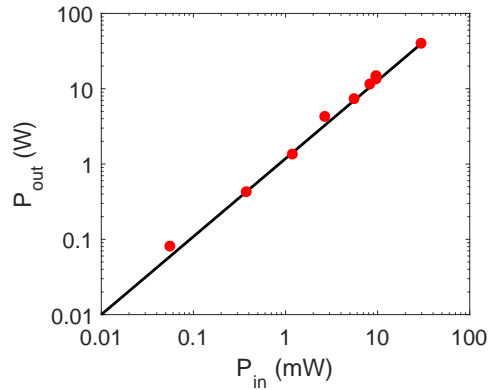}
	\caption {The linearity of the amplifier measured (red dots) at 251 GHz by varying the input power. The black line is a linear fit to the measured data at 251~GHz for the parameters in \figref{fig:gainbw} for Operating Point 1.}
  \label{fig:lineargain}
\end{figure}

\begin{figure}[!t]
  \centering
  \includegraphics[width=0.38\textwidth]{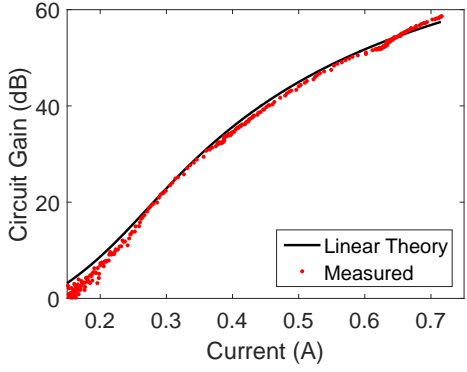}
	\caption {The circuit gain of the amplifier measured at 249 GHz as a function of the electron beam current and compared to the theoretical linear gain for the parameters in \figref{fig:gainbw} for Operating Point 1.}
  \label{fig:circuitgain}
\end{figure}

During operation the performance of the amplifier was optimized to achieve the highest gain-bandwidth product. The measured gain-bandwidth is shown in \figref{fig:gainbw} for the nominal operating conditions of 22.8 kV and 675 mA. With a bandwidth of 8 GHz and a peak device gain of 38 dB this represents a factor of 20 improvement in the gain-bandwidth from our previous measurements. The beam power achievable with zero drive stable operation also increased a factor of two, indicating that the increased circuit length is aiding with the suppression of oscillations in the uptaper sections. \figref{fig:gainbw} includes a second higher voltage operating point which was also utilized for collecting short pulse amplification data. The linearity of the amplifier was demonstrated experimentally as shown in \figref{fig:lineargain}. Saturation behavior was not observed due to insufficient input power from the solid-state driver source. A comparison between the measured and theoretical circuit gain from linear theory is shown in \figref{fig:circuitgain}. The measured circuit gain is calculated by adding to the measured device gain both the input/output coupling loss measured during cold test\cite{nanni2013250} ($\sim$11~dB) and the calculated insertion loss ($\sim$9~dB) that results from the coupling of the input signal into three different modes that interact with the electron beam of which only one is amplified.\cite{nusinovich1992theory,nanni2013250} \figref{fig:circuitgain} indicates that further improvement to the millimeter-wave transport and coupling into and out of the circuit can provide a significant increase in overall device gain. 

\begin{figure}[!t]
  \centering
  \includegraphics[width=0.45\textwidth]{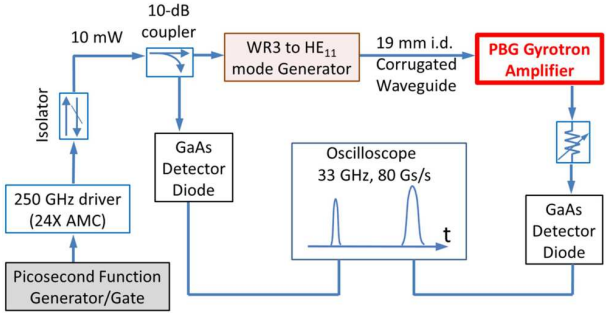}
	\caption {Schematic of the short pulse amplification measurement setup in a Photonic-Band-Gap gyrotron amplifier.}
  \label{fig:schematic}
\end{figure}

\begin{figure}[!t]
  \centering
  \includegraphics[width=0.38\textwidth]{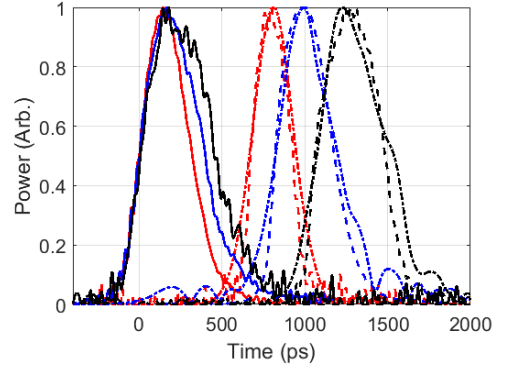}
	\caption {Input and output pulses measured with a variety of input pulse widths at a center frequency of 249.7 GHz. Pulse compression is observed when the bandwidth of the amplified pulse is increased due to the gain-bandwidth for the parameters in \figref{fig:gainbw} for Operating Point 2. The traces with solid lines (--) are for measured input pulses, the dashed lines (-~-) are for measured output pulses, and the dash-dotted lines (-$\cdot$) are for the modeled output pulse using the input pulse waveform. For the red traces the measured input pulse width was 320~ps with a measured/modeled output pulse width of 260/285~ps, for the blue traces the measured input pulse width was 38~ps with a measured/modeled output pulse width of 360/380~ps, and for the black traces the measured input pulse width was 440~ps with a measured/modeled output pulse width of 425/460~ps.}
  \label{fig:compression}
\end{figure}

The microwave source for the amplifier is a VDI 24X amplifier-multiplier chain that produces 30~mW over a 10~GHz bandwidth. By gating the local oscillator which is fed to the amplifier-multiplier chain we produce pulses as short as 250~ps that are sent on low-loss, low-dispersion highly-overmoded corrugated waveguide to the input of the gyrotron amplifier. 
To measure the pulse waveform two calibrated GaAs detector diodes were used at the input and output of the gyrotron amplifier and the waveform was recorded on a Keysight Infiniium 90000 X-Series oscilloscope with 33~GHz bandwidth and 80~Gs/s data collection rate. A schematic of the measurement setup is shown in \figref{fig:schematic}. An example of the measured temporal profile of the input and output picosecond pulses using this setup at 249.7~GHz is shown in \figref{fig:compression} for three pulse widths. For the longest input pulse width (black) the input and output pulse lengths are nearly identical at 440/425~ps, respectively. As the input pulse width is reduced we observe an output pulse which is compressed with respect to the input. This pulse compression can be attributed to the broadening of the pulse bandwidth. 

In order to model the behavior of the time evolution and amplification of the picosecond long pulses, we employ the modified dispersion relation in the presence of an electron beam in the small signal regime where the amplified mode propagates as exp$(ik(\Gamma-\bar \Delta)z)$. The coupled dispersion relation 
given by

\begin{equation}
{\Gamma ^3} - \bar \Delta {\Gamma ^2} - I_{0}^{'}(n - b)\Gamma  + (\bar \mu  - \bar \Delta b)I_{0}^{'} = 0,
\label{eq:modifiedDR}
\end{equation} 
  
where, $b=({k_z}/k)\beta _{\bot}^2/2{\beta _{z}}(1 - ({k_z}/k){\beta _{z}})$, $\bar \Delta=(1/{\beta _{z}}) - (n\Omega /{\beta _{z}}{\gamma}\omega ) - ({k_z}/k)$, $\bar \mu=(\beta _{ \bot}^2/2{\beta_{z}})(1 - {({k_z}/k)^2})/(1 - ({k_z}/k){\beta _{z}})$, $\beta=v/c$, $\beta_{\bot}=v_\perp/c$, $k$ is the wavenumber, $n$ is the cyclotron harmonic. $I_{0}^{'}$ is the normalized current given by Nusinovich and Li.\cite{nusinovich1992theory} 

It was shown previously \cite{kim2010prl} that the pulse interaction with the electron beam can be described by
\begin{equation}
\frac{\partial}{\partial z}a(z,\omega)=-ik_z a - ik\Gamma a
\label{eq:prop}
\end{equation}
where $a(z,\omega)$ is the frequency spectrum of the pulse and $-ik\Gamma$  was replaced with the measured gain-bandwidth $G(\omega)$ shown as in \figref{fig:gainbw}. The effects of dispersion can be modeled as distortions to the frequency spectrum of the pulse as it travels through the interaction circuit. For a pulse $A(z_0,t)$ at the start of the circuit, the frequency spectrum is given by,
\begin{equation}
a(\omega)= \frac{1}{\sqrt{2\pi}}\int_{-\infty}^\infty \! A(z_0,t)e^{-i\omega t} \ dt .
\end{equation}
For the Fourier transform of a pulse, if we amplify and advance the phase of every spectral component, it will be equivalent to propagating the pulse through the circuit. Performing an inverse Fourier transform on this new spectrum
\begin{equation}
A'(z_1,t)= \frac{1}{\sqrt{2\pi}}\int_{-\infty}^\infty \! a(\omega)e^{-ik_z z_1}e^{G(\omega)z_1}e^{i\omega t} \, d\omega
\label{eq:IFTnewpulse}
\end{equation}
produces the output pulse in the time domain. Three examples of the simulated output pulse using the above explanation are shown in \figref{fig:compression} along with the measured input and output pulses. The measured temporal profile of the input pulse is used in numerical simulation to generate the output pulse. An excellent agreement is seen between the measured and modeled output profiles.

\begin{figure}[!t]
  \centering
  \includegraphics[width=0.38\textwidth]{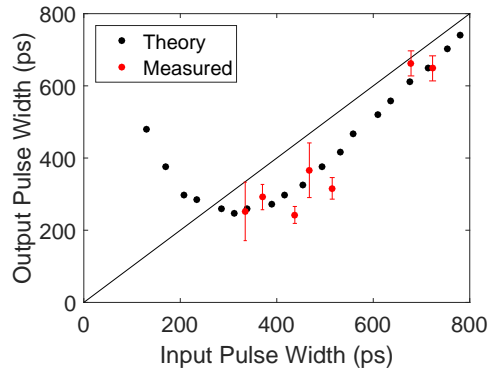}
	\caption {Comparison between pulse width measured and numerically simulated at the input and output of the amplifier with a center frequency of 251.4 GHz. A solid black line is shown for equal pulse widths for the parameters in \figref{fig:gainbw} for Operating Point 1.}
  \label{fig:pulseinout}
\end{figure}

To investigate the behavior of short-pulse amplification for the wide gain-bandwidth spectrum as observed in \figref{fig:gainbw} both measurements and numerical calculations were performed for input pulses of Gaussian shape of various temporal lengths. Calculations were performed at 251.4 GHz, maintaining a frequency far from the circuit cutoff frequency ($\sim$235~GHz), for the operating conditions listed in \figref{fig:gainbw}. The wavenumber $k_{z}+k\Gamma$ used in calculations in Eq.~(\ref{eq:prop}) is obtained from the coupled dispersion relation in Eq.~(\ref{eq:modifiedDR}). \figref{fig:pulseinout} shows the numerical calculations of the input vs output pulse width for the gyrotron amplifier. Theoretical calculations were performed with the best fitted Gaussian shape to the measured input pulse at a frequency of 251.4~GHz. \figref{fig:pulseinout} shows that as the input pulse width decreases from $\sim$800~ps, the output pulse is compressed, with a minimum output pulse width of $\sim$250~ps achieved for an input pulse of $\sim$300~ps. At the shortest input pulse widths, below $\sim$250~ps, the output pulse shows broadening. 

Along with the numerical calculations, pulse widths were also observed experimentally as shown in \figref{fig:pulseinout}. We observe compression for input pulse widths of 335~ps to output pulse widths of 250~ps, agreeing well with theory. We do not observe any nonlinear distortion to the shape of the pulse as we are still operating in the linear regime of the amplifier. The effect of pulse compression can be attributed to the fact that the operating frequency was chosen far, $>$~10 GHz, from the cutoff frequency and the spectral bandwidth of the amplified pulse is small compared with the gain-bandwidth of the amplifier. These factors combine to increase the bandwidth of the pulse while avoiding a significant amount of dispersion which increases the pulse width. Comparing results from \figref{fig:compression} and \figref{fig:pulseinout} we observe that the achievable compression increases when the center frequency of the input pulse is both further from cutoff, and it also offset higher in frequency from the center of the gain spectrum. In addition, the slightly wider bandwidth of Operating Point 1 results in a narrower achievable pulse width. These results are unlike our earlier experiment \cite{kim2010prl} where pulse broadening was observed with a narrow gain-bandwidth spectrum while operating near cutoff thus increasing the waveguide dispersion contribution to the broadening.    

In summary, we have observed more than 55~dB of circuit gain and a very wide 16~GHz of operational bandwidth at 250 GHz using a photonic-band-gap gyrotron amplifier. The amplifier operated in the linear regime and the output power was limited due to input power from the solid-state driver. Amplification of picosecond long pulses was observed without any distortion to the temporal shape of the pulse. We observed compression of the pulses shorter than 800~ps when the amplifier was operated far from the cutoff frequency and the effect of waveguide dispersion was minimized. This agrees well with numerical simulations using the coupled dispersion relation. 

The authors gratefully acknowledge Ivan Mastovsky for his assistance. This work was supported by the National Institutes of Health (NIH) and the National Institute for Biomedical Imaging and Bioengineering (NIBIB) under Grants No. EB001965 and No. EB004866. 

\bibliography{PhDbib}
\end{document}